\documentclass[11pt]{article}
\usepackage[a4paper,margin=1in]{geometry}
\usepackage[numbers,sort&compress]{natbib}
\usepackage{graphicx}
\DeclareGraphicsExtensions{.png,.tif}
\usepackage{amsmath,amssymb}
\usepackage{booktabs}
\usepackage{tabularx}
\newcommand{\eqbox}[1]{%
  \setbox0=\hbox{$\displaystyle #1$}%
  \ifdim\wd0>\textwidth
    \resizebox{\textwidth}{!}{$\displaystyle #1$}%
  \else
    \box0%
  \fi}
\newcommand{\keywords}[1]{\par\noindent\textbf{Keywords: } #1\par}
\begin{document}

\title{Dynamical closure, stability, and the three-stage cosmological evolution chain of two $f(R,\mathcal{G})$ models selected by generalized Noether symmetry}
\author{Han Dong$^{\ast}$ \\
\small School of Electronic Science and Technology, \\
Xiamen University Tan Kah Kee College, Zhangzhou, China \\
Key Laboratory of Intelligent Manufacturing Equipment \\
and Industrial Internet Technology, \\
Fujian Provincial Universities, Zhangzhou, China \\
$^{\ast}$Corresponding author: donghan@mail.nankai.edu.cn \\
ORCID: https://orcid.org/0000-0002-1528-7142}
\date{29 November 2026}
\maketitle

\begin{abstract}
We examine whether two generalized-Noether-selected models can realize radiation domination, a standard matter era, and late-time acceleration. The models are analysed separately. For the polynomial model, promoting the Ricci-scalar time derivative yields a closed four-dimensional autonomous system. Its Jacobian spectrum is derived and checked numerically, and its field-equation consistency is verified. Heteroclinic orbits are integrated from radiation initial data and near the matter saddle. The accelerating end state is a transversally stable attractor, but none of 18 radiation trajectories reaches it; 12 reach the vanishing Gauss--Bonnet boundary, and all matter-era durations vanish. Even near the matter saddle, the longest matter-dominated interval is 1.56 e-folds, far below the roughly 8.5 required for structure formation. Within this closure, divergence at the vanishing invariant separates radiation from acceleration. For the product model, the Noether exponent relation is inserted into a published reduced system and the resulting eigenvalues are checked numerically. Stable acceleration occurs only in two unbounded ranges, while radiation evolves directly into acceleration without a matter era. Neither model supports the full chain in the formulation analysed: the polynomial result depends on the chosen closure and the product result is a formal extrapolation, not a strict verdict on the original model.

\end{abstract}


\section{Introduction}
\label{sec:intro}

\subsection{Observational establishment of late-time acceleration and the fine-tuning problem of the cosmological constant}
\label{sec:intro-obs}

Type~Ia supernovae provided direct evidence for late-time cosmic acceleration \citep{riess1998observational,perlmutter1999measurements}. WMAP9 \citep{hinshaw2013nine}, Planck~2016 \citep{ade2016planck}, and Planck~2018 \citep{aghanim2020planck} support a nearly flat Universe dominated by a negative-pressure component, while BAO measurements from 2dFGRS and SDSS \citep{cole20052dfgrs,eisenstein2005detection} and their joint analyses \citep{aubourg2015cosmological,carvalho2016constraints} give complementary distance--redshift constraints. High-redshift galaxy ages and age--redshift relations \citep{krauss1995age,alcaniz1999cosmological,alcaniz2001constraining} disfavour a decelerating matter-only Universe. Together these probes support a history of radiation domination, standard matter domination, and renewed acceleration. Identifying the late-time component with a cosmological constant is compatible with the data but faces the fine-tuning problem \citep{weinberg1989cosmological}: quantum-field-theoretic vacuum-energy estimates exceed the effective value required for acceleration by many orders of magnitude, and the present matter--dark-energy coincidence lacks a dynamical explanation. This motivates searching for a geometric or field-theoretic origin that can evolve by itself.

\subsection{Theoretical motivation and cost of Gauss--Bonnet modified gravity}
\label{sec:intro-gb}

Rewriting gravity on the geometric side is a common route to these difficulties; the representative step replaces the $R$ in the action by a general function $f(R)$, whose forms, field equations, and cosmological solutions have been systematically reviewed \citep{defelice2010f}. When higher-order curvature combinations are considered, the Gauss--Bonnet invariant $\mathcal{G}=R^2-4R_{\mu\nu}R^{\mu\nu}+R_{\mu\nu\rho\sigma}R^{\mu\nu\rho\sigma}$ occupies a special position: it appears in the renormalization and trace anomaly of quantum field theory in curved spacetime \citep{birrell1982quantum}, and in four dimensions it is a topological term, so that adding it linearly does not modify the field equations; nontrivial dynamics arises only through a nonlinear function $f(\mathcal{G})$, $f(R,\mathcal{G})$, or a coupling to a scalar field. Dark-energy models and cosmological solutions built in this way have been proposed since the beginning of this century \citep{nojiri2005gauss,nojiri2005dark}, and were later extended to discussions of late-time acceleration and phase-space structure \citep{li2007cosmology,defelice2009solar,ivanov2012dynamics,alimohammadi2009phase,cognola2006dark}. Higher-order derivatives bring a cost: Ostrogradsky-type ghost degrees of freedom, the behavior of tensor and scalar perturbations, and the stability of the additional modes all require separate scrutiny \citep{defelice2010matter,defelice2010final,calcagni2006ghost}, and stability analyses of the de~Sitter and key fixed points have become a standard step in judging whether a model can support a cosmological evolution \citep{delaCruzDombriz2012cosmological}. The choice of the function is thereby placed before the dynamical calculation rather than treated as an arbitrary starting point.

\subsection{Model selection by Noether symmetry and the surviving exponent ambiguity}
\label{sec:intro-noether}

Symmetry provides a systematic way to constrain the functional form: Noether identities, grounded in the application of Lie groups to differential equations \citep{olver1993applications}, have been used to select gravitational models \citep{capozziello2014noether,bajardi2023dynamical,dong2013new,capozziello2016noether}. Applying a generalized Noether theorem to Gauss--Bonnet cosmology \citep{dong2019noether} gives $n=2m$ for the additive ansatz $f=k_1R^n+k_2\mathcal{G}^m$ (Eq.~(35) of Ref.~\citep{dong2019noether}, Model~1) and $\frac n2+m-1=0$ for the product ansatz $f=kR^n\mathcal{G}^m$ (Eq.~(67) of Ref.~\citep{dong2019noether}, Model~2). Both retain only time translation, have vanishing conserved quantities, and leave $n$ undetermined \citep{dong2019noether}. The separate $m=1-n$ branch was judged unsuitable for accelerated expansion \citep{capozziello2014noether}; that verdict does not apply to Model~2.

\subsection{Dynamical systems analysis and the closure requirement}
\label{sec:intro-dsa}

Dynamical systems analysis (DSA) is the standard framework for testing cosmological evolutions: building on the treatment of Amendola et al.\ \citep{amendola2007conditions} and subsequent developments of the phase-space method \citep{carloni2005dynamical,carloni2015dynamical}, the Friedmann equations are cast as a dimensionless first-order autonomous system, fixed points are identified with cosmological stages, and the eigenvalues of the linearized Jacobian classify attractors, repellers, or saddles. The framework carries an implicit closure requirement: higher-order quantities such as $\Gamma=\ddot f_R/(f_RH^2)$ must be expressible explicitly in terms of the chosen variables, otherwise the system retains a free function and neither fixed points nor their stability can be determined \citep{amendola2007conditions}. Santos~da~Costa et al.\ enumerated eight fixed points for $f=\alpha R^n\mathcal{G}^{1-n}$ (the $m=1-n$ branch) and concluded that no standard matter-dominated era exists and no complete trajectory from radiation to acceleration can be formed \citep{dacosta2018dynamical}. Singh performed a parallel fixed-point classification for $f_0R^\alpha\mathcal{G}^{1-\alpha}$ and $f_0R^n+f_1\mathcal{G}$, listing the results in four tables with no phase-space trajectory, and deferred observational tests to future work \citep{singh2025dynamical}. The exponent relations adopted in these external analyses are the $m=1-n$ branch and the $m=1$ special case, both different from the $m=n/2$ and $m=1-n/2$ considered here \citep{dacosta2018dynamical,singh2025dynamical}; the fixed-point structure is highly sensitive to the exponents, so transferring these conclusions to Model~1 (polynomial type, $m=n/2$) or Model~2 (product type, $m=1-n/2$) lacks justification \citep{dong2019noether,dacosta2018dynamical,singh2025dynamical}, and a model-by-model reanalysis cannot be omitted.

\subsection{Research gap and contributions of this work}
\label{sec:intro-gap}

Four gaps motivate the analysis. The exponent relations of Models~1 and 2 are often conflated with the $m=1-n$ branch and the $m=1$ special case \citep{dong2019noether,dacosta2018dynamical,singh2025dynamical}; the additive-separable form does not close because $\Gamma$ reduces to an identity, so a fixed-point-level no-go cannot decide the trajectory-level chain \citep{amendola2007conditions}; previous classifications lack a quantitative comparison between matter-era residence times and the $\Delta N\approx8.5$ e-folds required by structure formation \citep{amendola2007accelerating,delaurentis2015cosmological}; and Model~2 lacks a parameter-region result after the Noether relation is inserted \citep{dong2019noether,dacosta2018dynamical}.

We therefore treat the two models separately. For Model~1, $\dot R$ is promoted to an independent variable, a closed four-dimensional system is subjected to a Bianchi consistency test, and the resulting explicit spectrum and saddle family are tested against radiation-initialized and saddle-neighbourhood orbits. For Model~2, $m=1-n/2$ is inserted into the reduced system of Ref.~\citep{dacosta2018dynamical} as a formal extrapolation checked by direct numerical Jacobians. The Model~1 verdict belongs to the closed system; the Model~2 verdict is confined to the reduced system and is not a strict dynamical result for Eq.~(67) of Ref.~\citep{dong2019noether}. No observational fit is claimed; only reproducible dynamical criteria are given.

The remainder of the paper is organized as follows. Section~\ref{sec:model} presents the common theoretical framework, the origin of the two models, and the construction of the closed system and of the reduced extrapolation. Section~\ref{sec:results} reports the fixed-point spectrum, the saddle screening, and the numerical orbits of Model~1, followed by the parameter regions and the fixed-point spectrum of Model~2. Section~\ref{sec:discussion} summarizes the stage-by-stage verdicts and discusses the methodological boundaries. Section~\ref{sec:conclusion} concludes.

\section{Methods and models}
\label{sec:model}

\subsection{Common theoretical framework}
\label{sec:model-framework}

Both models share the same field equations of $f(R,\mathcal{G})$ gravity and the same set of dimensionless variables. We adopt the form of the action with the matter term outside the gravitational coupling factor \citep{dacosta2018dynamical,defelice2010f},
\begin{equation}
\mathcal{S}=\int d^4x\,\sqrt{-g}\left[\frac{f(R,\mathcal{G})}{2\kappa^2}+\mathcal{L}_m\right],\qquad \kappa^2=8\pi G_N,
\label{eq:action}
\end{equation}
where the Gauss--Bonnet invariant is $\mathcal{G}=R^2-4R_{\mu\nu}R^{\mu\nu}+R_{\mu\nu\rho\sigma}R^{\mu\nu\rho\sigma}$ \citep{birrell1982quantum}. In a spatially flat Friedmann--Lema\^{\i}tre--Robertson--Walker metric, $R=6(2H^2+\dot H)$, $\mathcal{G}=24H^2(H^2+\dot H)$, and we use the logarithmic time $N=\ln a$ with $dN=H\,dt$ \citep{dacosta2018dynamical}. The two independent Friedmann equations read
\begin{equation}
\eqbox{3f_RH^2=\kappa^2(\rho_m+\rho_r)+\frac12\left(f_RR-f-6H\dot f_R+\mathcal{G}f_{\mathcal{G}}-24H^3\dot f_{\mathcal{G}}\right),}
\label{eq:friedmann1}
\end{equation}
\begin{equation}
\eqbox{2f_R\dot H=-\kappa^2\left(\rho_m+\frac43\rho_r\right)+H\dot f_R-\ddot f_R+4H^3\dot f_{\mathcal{G}}-8H\dot H\,\dot f_{\mathcal{G}}-4H^2\ddot f_{\mathcal{G}}.}
\label{eq:friedmann2}
\end{equation}

Following the variable choice of Santos~da~Costa et al.\ \citep{dacosta2018dynamical} we define the seven dimensionless quantities
\begin{equation}
\eqbox{\begin{aligned}
x_1&=\frac{\dot f_R}{f_RH},\quad x_2=\frac{f}{6f_RH^2},\quad x_3=\frac{R}{6H^2},\quad x_4=\frac{\kappa^2\rho_r}{3f_RH^2},\\
x_5&=\frac{\mathcal{G}f_{\mathcal{G}}}{6f_RH^2},\quad x_6=\frac{4H\dot f_{\mathcal{G}}}{f_R},\quad x_7=\frac{\kappa^2\rho_m}{3f_RH^2}
\end{aligned}}
\label{eq:vars}
\end{equation}
The first Friedmann equation then yields the algebraic constraint
\begin{equation}
1=-x_1-x_2+x_3+x_4+x_5-x_6+x_7,
\label{eq:constraint}
\end{equation}
with $\Omega_m=x_7$, $\Omega_r=x_4$, and $\Omega_{GDE}=1-x_4-x_7$. The effective equation-of-state parameter and the deceleration parameter are $\omega_{eff}=-\frac13(2x_3-1)$ and $q=1-x_3$. Radiation domination, standard matter domination, and de~Sitter acceleration therefore correspond to $x_3=0$, $x_3=1/2$, and $x_3=2$, respectively.

The general procedure for classifying the stability of a fixed point is to linearize the autonomous system at the equilibrium and read off the signs of the real parts of the nonzero Jacobian eigenvalues: all negative for an attractor, all positive for a repeller, and mixed signs for a saddle; eigenvalues with zero real part correspond to neutral directions in phase space and are not counted in the classification \citep{dacosta2018dynamical,amendola2007conditions,carloni2005dynamical}. This criterion applies equally to both models; what differs is only how the theoretical closure quantities $b=Rf_{RR}/f_R$ and $\Gamma=\ddot f_R/(f_RH^2)$ are expressed explicitly in terms of the chosen variables.

\subsection{Origin and exponent relations of the two models}
\label{sec:model-origin}

On the de~Sitter background $a\propto e^{H_0t}$, where $R=-12H_0^2$, $\mathcal{G}=24H_0^4$, and $\dot R=\dot{\mathcal{G}}=0$, the time-independence condition of the conserved quantity and the energy condition give closed constraints on the two ans\"atze \citep{dong2019noether}. For the additive-separable type these are $n=2m$ and $-k_1(-12)^n=k_2(24)^{n/2}$, yielding Model~1 (Eq.~(35) of Ref.~\citep{dong2019noether}),
\begin{equation}
f(R,\mathcal{G})=k_1R^n+(6)^{n/2}(-1)^{n+1}k_1\mathcal{G}^{n/2},\qquad m=\frac n2.
\label{eq:model1}
\end{equation}
For the product type, the time-independence condition of the conserved quantity $I=a^3kR^n\mathcal{G}^m\left(\frac n2+m-1\right)$ gives $\frac n2+m-1=0$, yielding what we call Model~2 (Eq.~(67) of Ref.~\citep{dong2019noether}),
\begin{equation}
f(R,\mathcal{G})=k\left(\frac{R}{\sqrt{\mathcal{G}}}\right)^n\mathcal{G}=kR^n\mathcal{G}^{1-n/2},\qquad m=1-\frac n2.
\label{eq:model2}
\end{equation}

Both models retain only $X=\partial_t$, have vanishing conserved quantities, and leave $n$ open \citep{dong2019noether}. The source work also discussed the $m=1-n$ branch, whose power-law solutions were judged unsuitable for acceleration \citep{dong2019noether,capozziello2014noether}; it is not our Model~2.

The methodological distinction is equally important. Model~1 has $f_{R\mathcal{G}}=0$ and is treated by the DSA framework of Ref.~\citep{dacosta2018dynamical} with the $\Gamma$-closure idea of Ref.~\citep{amendola2007conditions}: promoting $\dot R$ gives an explicit spectrum, saddle screening, and orbit tests, reported in Figs.~\ref{fig:model1-evolution} and \ref{fig:model1-phase} and Tables~\ref{tab:fp}--\ref{tab:neighbourhood}. Model~2 has $f_{R\mathcal{G}}\neq0$ and is examined by inserting $m=1-n/2$ into the reduced system of Ref.~\citep{dacosta2018dynamical}, combining $\lambda_{x_5}$, numerical Jacobians, and fixed-point spectra, with results in Figs.~\ref{fig:model2-stability} and \ref{fig:model2-evolution} and Tables~\ref{tab:lambda}--\ref{tab:stability}. The former conclusions are strict within the closed system; the latter are a formal extrapolation within the reduced system. The $m=1-n$ branch of Ref.~\citep{dacosta2018dynamical} and the $m=1$ case of Ref.~\citep{singh2025dynamical} are external reference forms only, intersecting Model~2 at $n=0$ and Model~1 at $n=2$, respectively.

\subsection{Model 1: construction of the closed autonomous system}
\label{sec:model-closure}

Model~1 has $f_R=nk_1R^{n-1}$, $f_{RR}=n(n-1)k_1R^{n-2}$, $f_{R\mathcal{G}}=0$, hence $b=n-1$. Substituting $\ddot f_R=f_{RRR}\dot R^2+f_{RR}\ddot R$ directly into the definition of $\Gamma$ gives
\begin{equation}
\Gamma=\frac{dx_1}{dN}+x_1^2+x_1(x_3-2),
\label{eq:gamma-identity}
\end{equation}
which is merely the inverse of the first equation of the system---an identity, not a theoretical input \citep{amendola2007conditions}. Without supplementary information $\Gamma$ therefore cannot be expressed through the seven variables of Eq.~\eqref{eq:vars}, and the original reduced system does not close on Model~1; the notion of a ``four-equation autonomous system'' is untenable as it stands.

The closure adopted here promotes $\dot R$ to an independent variable. Let
\begin{equation}
y\equiv\frac{\dot R}{6H^3},\qquad x_1=(n-1)\frac{y}{x_3},\qquad \delta\equiv\frac{\dot H}{H^2}=x_3-2.
\label{eq:y-def}
\end{equation}
From $R=6(2H^2+\dot H)$ one obtains $\ddot R=6H^4(3\delta y+y')$, so that $\ddot f_R$ depends linearly on $y'\equiv dy/dN$, and the second Friedmann equation degenerates into a linear algebraic equation for $y'$. Writing $Q=y+2\delta^2$ and $\Delta=x_3-1$, the constraint yields
\begin{equation}
\eqbox{x_5=\frac{1+x_1+x_3/n-x_3-x_4-x_7}{1-2/n-(m-1)Q/\Delta^2},\qquad C=\frac{(m-1)x_5}{\Delta^2},\qquad D=\frac{n-1}{x_3}+C,}
\label{eq:x5}
\end{equation}
where $x_2=x_3/n+x_5/m$ and $x_6=CQ$. The second-derivative terms become
\begin{equation}
\eqbox{\frac{4\ddot f_{\mathcal{G}}}{f_R}=C\left[2\delta^3-16\delta^2+9\delta y+y'+\frac{(m-2)Q^2}{\Delta}\right],\quad \Gamma=\frac{n-1}{x_3}(3\delta y+y')+\frac{(n-1)(n-2)y^2}{x_3^2}.}
\label{eq:second-derivs}
\end{equation}
Substituting these into the second Friedmann equation and dividing by $f_RH^2$, one solves for $y'=(A_0-2\delta)/D$, where
\begin{equation}
\eqbox{\begin{aligned} A_0=&-3x_7-4x_4+x_1-\frac{3\delta y(n-1)}{x_3}-\frac{(n-1)(n-2)y^2}{x_3^2}\\ &+CQ(1-2\delta)-C\left[2\delta^3-16\delta^2+9\delta y+\frac{(m-2)Q^2}{\Delta}\right]. \end{aligned}}
\label{eq:A0}
\end{equation}
With the state vector $\mathbf{s}=(x_3,x_4,x_7,y)$ this gives the closed four-dimensional autonomous system
\begin{equation}
\eqbox{x_3'=y-2x_3\delta,\qquad x_4'=-x_4(2x_3+x_1),\qquad x_7'=-x_7(2x_3+x_1-1),\qquad y'=\frac{A_0-2\delta}{D}.}
\label{eq:closed-system}
\end{equation}
Since $y'$ is now determined by the field equations, $\Gamma$ becomes an explicit function of the state, which is fully equivalent to the $\Gamma(x_i)$ closure required by Amendola et al., differing only in whether the fourth degree of freedom is denoted by $y$ or by $\Gamma$ \citep{amendola2007conditions}. Differentiating the constrained $x_5$ along the closed system and comparing it with the general $x_5$ evolution equation obtained before the reduction of Ref.~\citep{dacosta2018dynamical}, the residuals at generic states with $n=0.75,1.5,2.5$ lie between $3.8\times10^{-12}$ and $5.1\times10^{-10}$. This checks consistency with the field equations before the reduction, not with the later reduced third equation.

Numerical integration must avoid four classes of singularities: $x_3=0$, where $R=0$ and $f_R\propto R^{n-1}$ degenerates the variable set; $x_3=1$, where $\Delta=0$, $\mathcal{G}=0$, and both $Q^2/\Delta$ and $f_{\mathcal{G}}\propto\mathcal{G}^{m-1}$ diverge; $D=0$, where the denominator of $y'$ vanishes; and $n=2$, where $m=1$ and $x_5$ cannot be determined from the constraint. These singularities constitute the topological boundary for the trajectory verdicts below.

\subsection{Model 2: formal extrapolation of the reduced system}
\label{sec:model-reduced}

In deriving the reduced system of Eqs.~(38) and (39) of Ref.~\citep{dacosta2018dynamical}, Santos~da~Costa et al.\ kept $m$ as a general parameter and substituted $m=1-n$ only in the step of solving for the fixed points \citep{dacosta2018dynamical}. Model~2 belongs to the same form $f=\alpha R^n\mathcal{G}^m$, so replacing $m$ by $1-n/2$ in that reduced system is a legal substitution at the level of this reduced system; $b=n-1$ is unchanged. The reduced system takes $(x_3,x_4,x_5,x_7)$ as independent variables, with $x_1$ given algebraically by Eq.~(39) of Ref.~\citep{dacosta2018dynamical}.

On the invariant submanifold $x_5=0$, the $(x_3,x_4,x_7)$ subsystem is independent of $m$, and the third row of the $4\!\times\!4$ Jacobian has only the $(3,3)$ element nonzero at $x_5=0$. The fourth eigenvalue can therefore be written in closed form,
\begin{equation}
\eqbox{\lambda_{x_5}=\frac{m}{x_3-1}\left[\frac{x_1x_3}{b}+2(x_3-2)^2\right]+\frac{x_1}{b}-2(x_3-2),\qquad b=n-1.}
\label{eq:lambda-x5}
\end{equation}
Substituting $m=1-n$ reproduces exactly the fourth eigenvalue obtained by strict linearization of Eqs.~(38) and (39) of Ref.~\citep{dacosta2018dynamical}: $8n-4$ for $P_3$, $(8n^2-12n+5)/(n-1)$ for $P_4$, and $(8n^2-12n+3)/(n-1)$ for $P_8$, confirming that this derivation is self-consistent with the reduced system. All conclusions are additionally cross-checked by constructing the full four-dimensional right-hand side with central differences and computing the numerical Jacobian eigenvalues directly, with point-by-point agreement.

This route carries an explicit boundary of applicability. The derivation of the reduced system of Eqs.~(38) and (39) of Ref.~\citep{dacosta2018dynamical} drops the cross term in $\dot f_R=f_{RR}\dot R+f_{R\mathcal{G}}\dot{\mathcal{G}}$, whereas the product type has $f_{R\mathcal{G}}\neq0$; the identity $x_5=mx_2=(m/n)x_3$ is not imposed, and the sign convention of Eq.~(36) of Ref.~\citep{dacosta2018dynamical} is also incompatible with the variable definitions \citep{dacosta2018dynamical}. The Model~2 results in Sect.~\ref{sec:results-model2} should therefore be understood only as a ``formal extrapolation within that reduced system'' and do not constitute a strict dynamical verdict on Eq.~(67) of Ref.~\citep{dong2019noether}.

\subsection{Numerical methods and criteria}
\label{sec:model-numerics}

The orbit integration for Model~1 employs the DOP853 solver \texttt{solve\_ivp} of \texttt{scipy} with relative tolerance $\texttt{rtol}=10^{-10}$, absolute tolerance $\texttt{atol}=10^{-12}$, maximum step size $0.02$, and an integration ceiling $N_{\max}=30$; the termination event is $\min(|x_3-1|,|x_3|,|D|,|\text{den}|)<2\times10^{-4}$. Radiation-dominated initial data are $x_3(0)=10^{-3}$, $x_4(0)=0.9989$, $x_7(0)=10^{-3}$, $y(0)=c\,x_3(0)$ with $c\in\{-2,0,2\}$; scanning $n\in\{0.75,\allowbreak 1.00,\allowbreak 1.25,\allowbreak 1.50,\allowbreak 1.75,\allowbreak 2.50\}$ gives 18 orbits in total.

The operational criterion for a standard matter-dominated segment is $|\omega_{eff}|<0.05$ together with $\Omega_m>0.8$; the longest continuous $N$-interval on an orbit satisfying these conditions is recorded as the matter-era e-folds. The observational benchmark is a matter era extending from $z_{\rm eq}\approx3400$ to the onset of acceleration at $z\sim0.7$, corresponding to $\Delta N\approx\ln(3400/0.7)\approx8.5$ e-folds \citep{delaurentis2015cosmological}; this duration is simultaneously the lower bound required by structure formation \citep{delaurentis2015cosmological}. Saddle-neighbourhood orbits start from the saddle of the $n=1$ continuous family and add perturbations $\epsilon\in\{10^{-4},\allowbreak 10^{-3},\allowbreak 10^{-2}\}$ to the state components, giving 54 orbits in total, which serve to place a quantitative upper bound on ``how long matter domination can be maintained even when the orbit is placed by hand near the matter point''.

The fixed points with $x_5\neq0$ and their stability in representative cases are obtained by multi-start numerical root finding, while the $P_2$ stability boundaries are derived analytically and checked against direct numerical Jacobians.

\section{Results}
\label{sec:results}

\subsection{Fixed-point structure of Model 1 and the explicitly $n$-dependent Jacobian spectrum}
\label{sec:results-model1-fp}

The equation $x_3'=0$ of the closed system gives $y=2x_3(x_3-2)$. Writing $p\equiv x_3$, numerical tests show that $y'$ vanishes to the resolution of the computation on $x_4=x_7=0$ (with $|y'|\lesssim10^{-12}$, covering $n=0.75,1.5,2.5$ and $p=0.2$--$3.0$), so the system admits a one-dimensional continuous family of fixed points,
\begin{equation}
\eqbox{\mathcal{M}_{\rm vac}:\quad (x_3,x_4,x_7,y)=(p,0,0,2p(p-2)),\qquad \omega_{eff}=\frac{1-2p}{3},\quad \Omega_{GDE}=1.}
\label{eq:mvac}
\end{equation}

Reordering the variables as $(x_3,y,x_4,x_7)$, the Jacobian becomes block lower triangular,
\begin{equation}
\eqbox{J\big|_{\mathcal{M}_{\rm vac}}=\begin{pmatrix}4(1-p) & 1 & 0 & 0\\ J_{41} & (4n-9)+(5-2n)p & \ast & \ast\\ 0 & 0 & 2n(2-p)-4 & 0\\ 0 & 0 & 0 & 2n(2-p)-3\end{pmatrix},\quad J_{41}=4(1-p)J_{44}}
\label{eq:jacobian}
\end{equation}
where the entries marked $\ast$ do not affect the spectrum. Since $y'\equiv0$ holds on the entire family, the determinant of the upper-left $2\times2$ block vanishes and one eigenvalue must be zero. This yields the closed-form spectrum in explicit dependence on $n$ and $p$,
\begin{equation}
\eqbox{\lambda\in\left\{\,0,\quad 2n(2-p)-4,\quad 2n(2-p)-3,\quad (4n-5)+(1-2n)p\,\right\},}
\label{eq:spectrum}
\end{equation}
whose four entries correspond respectively to the neutral direction along $\mathcal{M}_{\rm vac}$, the radiation perturbation, the matter perturbation, and the geometric perturbation. Using $n\in\{0.5,0.75,1.25,1.5,2.5,3.0\}$ and $p\in\{0.3,0.5,1.5,2.0,3.0\}$, 30 parameter sets in total, a central-difference Jacobian is compared with the closed form; the maximum deviation is $6.98\times10^{-5}$, limited by the difference step, confirming the closed form. The zero eigenvalue also explains the repeated zero roots in the existing DSA tables of the source manuscript \citep{dong2019noether}: they correspond to the neutral direction along the continuous family, not to additional isolated fixed points.

The de~Sitter point takes $p=2$, giving $\lambda=\{-4,-3,-3,0\}$, independent of $n$; the three transverse directions are all negative, so it is a transversally stable attractor for any $n$, with $\omega_{eff}=-1$ and $\Omega_{GDE}=1$. At $p=1/2$ ($\omega_{eff}=0$) the spectrum is $\{0,3n-4,3n-3,3n-4.5\}$. Table~\ref{tab:fp} summarizes the fixed-point families, spectra, and physical classification of the closed system.

\begin{table}[htbp]
\caption{Fixed-point structure and Jacobian spectra of the closed system of Model~1.}
\label{tab:fp}
\small
\begin{tabularx}{\textwidth}{@{}lXXX@{}}
\toprule
Object & State $(x_3,x_4,x_7,y)$ & Spectrum & Physical classification \\
\midrule
Vacuum family $\mathcal{M}_{\rm vac}$ & $(p,0,0,2p(p-2))$ & $\{0,\ 2n(2-p)-4,\ 2n(2-p)-3,\ (4n-5)+(1-2n)p\}$ & one-dimensional neutral direction, $\Omega_{GDE}=1$ \\
Standard radiation point $p=0$ & $(0,0,0,0)$ & $\{0,\ 4n-4,\ 4n-3,\ 4n-5\}$ & $\omega_{eff}=1/3$; an early repelling end for $n<1$ \\
Matter scaling $p=1/2$ & $(1/2,0,0,-3/2)$ & $\{0,\ 3n-4,\ 3n-3,\ 3n-4.5\}$ & $\omega_{eff}=0$ but $\Omega_m=0$ \\
de~Sitter point $p=2$ & $(2,0,0,0)$ & $\{-4,-3,-3,0\}$ & $\omega_{eff}=-1$; transversally attracting for any $n$ \\
$\Omega_m\neq0$ branch & $x_3=2-\frac{3}{2n}$ & see Eq.~\eqref{eq:saddle-x3} & $\omega_{eff}=1/n-1$ \\
\bottomrule
\end{tabularx}
\end{table}

\subsection{Saddle screening of Model 1}
\label{sec:results-model1-saddle}

If $\Omega_m=x_7\neq0$, then $x_7'=0$ requires $2x_3+x_1-1=0$; combined with $x_3'=0$, i.e.\ $x_1=2(n-1)(x_3-2)$, this gives
\begin{equation}
x_3=2-\frac{3}{2n},\qquad \omega_{eff}=\frac1n-1.
\label{eq:saddle-x3}
\end{equation}
Hence for $n\neq1$, $\omega_{eff}=0$ is strictly impossible, consistent with the fixed-point-level no-go obtained for the unclosed system. The only window is $n=1$: here $x_1=(n-1)y/x_3\to0$ remains finite, $\ddot f_R=0$, and after closure $n=1$ is no longer singular; at $x_3=1/2$ and $y=-3/2$, both $x_7'$ and $y'$ vanish identically for any $\Omega_m$, producing a continuous saddle family with $\omega_{eff}=0$ and free $\Omega_m$, with $\Omega_{GDE}=1-\Omega_m$. The constraint gives $x_5=(1-\Omega_m)/5$, and the closed denominator becomes
\begin{equation}
D=C=-\frac25(1-\Omega_m),
\label{eq:D-saddle}
\end{equation}
so that the pure-matter point $\Omega_m=1$ is a zero-over-zero limiting point of the closure and is not defined by the present formulation; the physical range considered is $0\le\Omega_m<1$.

Scanning $n\in\{0.60,\allowbreak 0.75,\allowbreak 0.90,\allowbreak 0.96,\allowbreak 0.98,\allowbreak 0.99,\allowbreak 1.01,\allowbreak \dots,\allowbreak 3.00\}$ and adding the $n=1$ continuous family, with the three screening criteria $|\omega_{eff}|<0.05$, $\Omega_m>0.8$, and a saddle Jacobian, yields exactly 3 strict matches, all belonging to the $n=1$ continuous family; no match occurs for $n\neq1$ (Table~\ref{tab:saddle}). The continuous family is non-hyperbolic at $\Omega_m=0$ and a saddle for every $\Omega_m>0$ in the physical range. As $\Omega_m\to1^-$, the positive eigenvalue diverges as $(1-\Omega_m)^{-1}$, showing that the closer to pure matter domination, the more unstable the saddle.

\begin{table}[htbp]
\caption{Strict matches of the Model~1 screening with $|\omega_{eff}|<0.05$, $\Omega_m>0.8$, and a saddle Jacobian.}
\label{tab:saddle}
\small
\begin{tabular}{@{}ccclc@{}}
\toprule
$n$ & $\Omega_m$ & $\Omega_{GDE}$ & Spectrum $\{0,\lambda_-,-1,\lambda_+\}$ & Type \\
\midrule
1.00 & 0.900 & 0.100 & $\{0,-7.50,-1,+6.00\}$ & saddle \\
1.00 & 0.950 & 0.050 & $\{0,-10.53,-1,+9.03\}$ & saddle \\
1.00 & 0.990 & 0.010 & $\{0,-23.01,-1,+21.51\}$ & saddle \\
\bottomrule
\end{tabular}
\end{table}

\subsection{Trajectory-level test of Model 1}
\label{sec:results-model1-orbits}

\begin{figure}[htbp]
\centering
\includegraphics[width=\columnwidth]{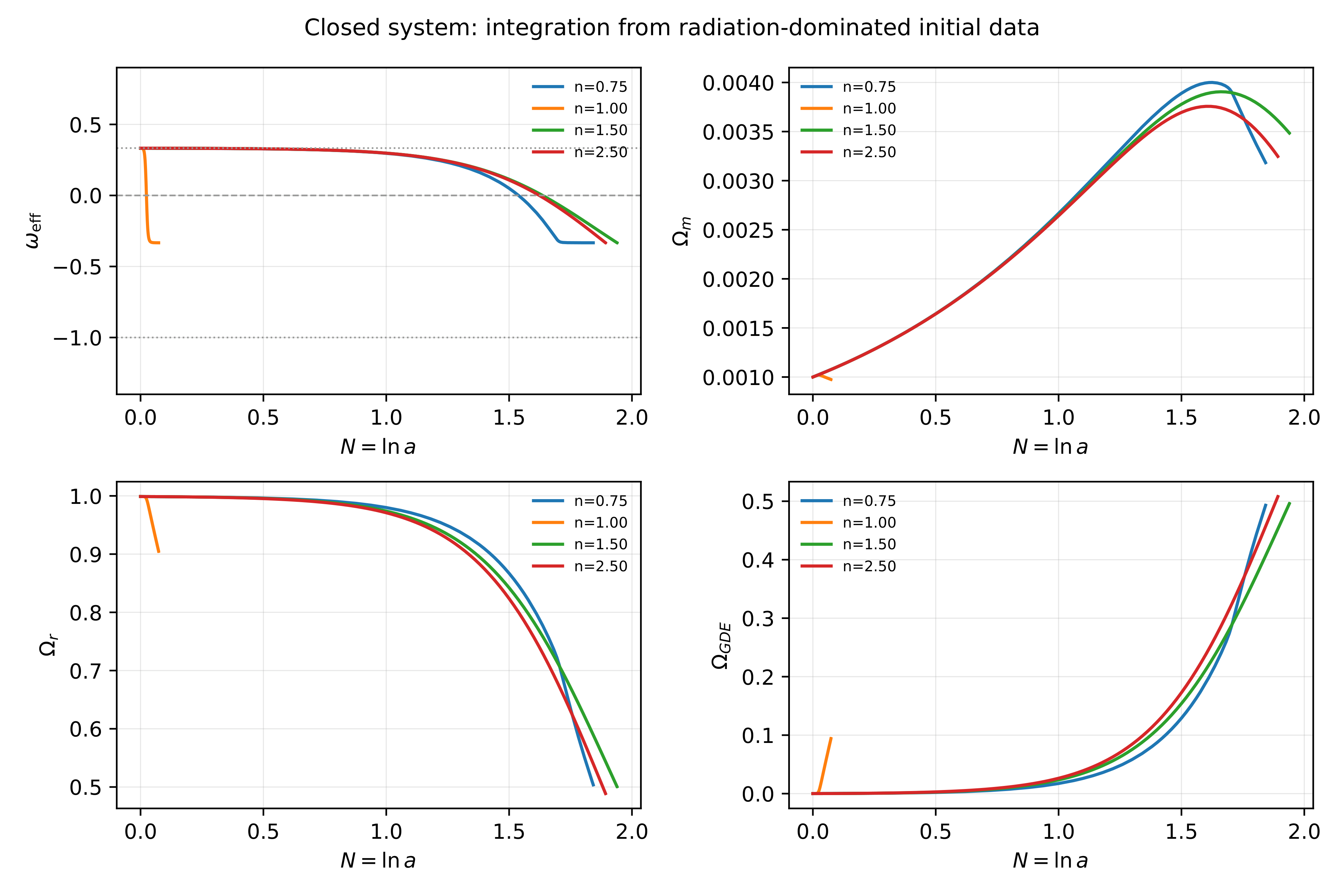}
\caption{Evolution of the closed Model~1 system from radiation-dominated initial data. Horizontal axis $N=\ln a$; the four panels show, from top to bottom, $\omega_{eff}$, $\Omega_m$, $\Omega_r$, and $\Omega_{GDE}$; curves correspond to $n=0.75,1.00,1.50,2.50$. All orbits terminate before reaching de~Sitter, $\Omega_m$ stays below $4\times10^{-3}$, and no standard matter-dominated segment appears.}
\label{fig:model1-evolution}
\end{figure}

Fixed-point stability alone cannot establish a heteroclinic chain; orbit integration is decisive. Figure~\ref{fig:model1-evolution} shows the radiation-initialized evolution for $n=0.75,1.00,1.50,2.50$ ($c=0$). All orbits begin at $\omega_{eff}\simeq1/3$, $\Omega_r\simeq1$, and terminate by $N\simeq1.5$--$1.9$. Matter never becomes dominant: $\Omega_m$ remains below about $0.004$, and $\omega_{eff}$ reaches only about $-0.4$, never the de~Sitter value $-1$. The $n=1.00$ orbit ends already at $N\simeq0.07$, where the $\mathcal{G}=0$ singularity is triggered prematurely.

\begin{figure}[htbp]
\centering
\includegraphics[width=\columnwidth]{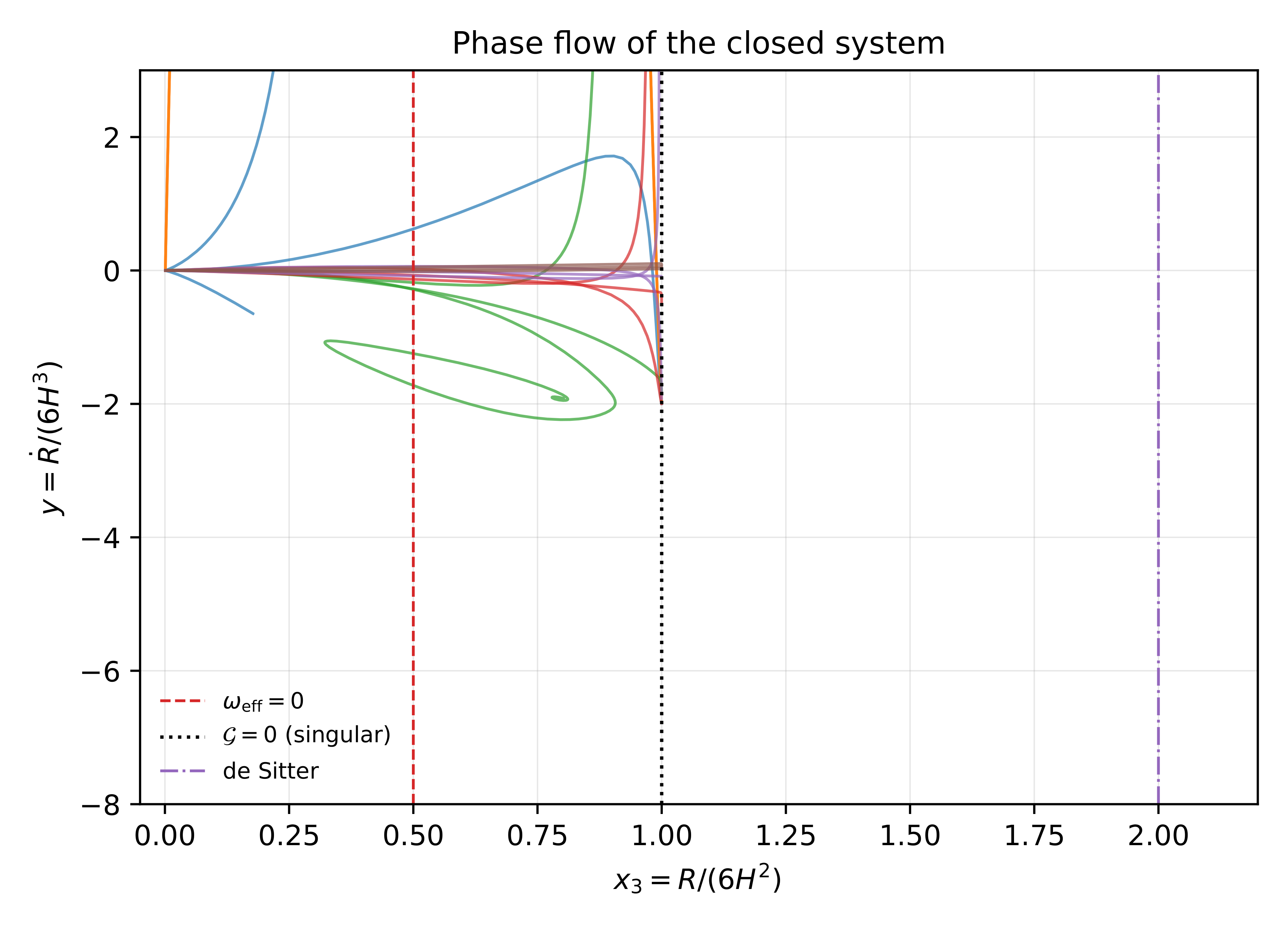}
\caption{Phase plane $(x_3,y)$ of the closed Model~1 system. The vertical lines mark, from left to right, $\omega_{eff}=0$ ($x_3=0.5$), the $\mathcal{G}=0$ singularity ($x_3=1$), and de~Sitter ($x_3=2$). All orbits are blocked by the singularity at $x_3=1$ and the region $x_3>1$ is inaccessible.}
\label{fig:model1-phase}
\end{figure}

The $(x_3,y)$ phase plane in Fig.~\ref{fig:model1-phase} shows the same obstruction directly. Every orbit starts near $x_3=0$, folds back or terminates at $x_3=1$, and none reaches the de~Sitter point $x_3=2$; the region $x_3>1$ remains empty. Since $\mathcal{G}=24H^4(x_3-1)$, the radiation side has $\mathcal{G}<0$ while de~Sitter has $\mathcal{G}>0$, so any continuous trajectory must cross $\mathcal{G}=0$, where $Q^2/\Delta$ and $f_{\mathcal{G}}\propto\mathcal{G}^{m-1}$ diverge. This topological barrier is intrinsic to the model and independent of initial data.

Table~\ref{tab:radiation-runs} lists the integration results of the 18 radiation-initialized orbits. Of the 18 orbits, 0 reach de~Sitter, 12 terminate at the combined event with $x_3=1$ ($|x_3-1|=2\times10^{-4}$), 3 terminate through $D=0$, 1 stops because the step size collapses near $D=0$, and 2 reach the integration ceiling with $x_3<1$; the matter-era e-folds vanish for all orbits. The $n=0.75, c=-2$ orbit is still at $x_3=0.177$ when $N_{\rm end}=30$ is reached, rather than entering the accelerating stage.

\begin{table}[htbp]
\caption{Integration results of the radiation-initialized orbits of Model~1 (representative entries).}
\label{tab:radiation-runs}
\small
\begin{tabular}{@{}ccllll@{}}
\toprule
$n$ & $c$ & $N_{\rm end}$ & $x_3(N_{\rm end})$ & Terminates at $\mathcal{G}=0$ & Matter-era e-folds \\
\midrule
0.75 & $-2$ & 30.00 & 0.177 & no & 0 \\
0.75 & $0$ & 1.84 & 0.9998 & yes & 0 \\
1.00 & $-2,0,2$ & 0.07 & 0.9998 & yes & 0 \\
1.25 & $0$ & 2.19 & 0.9998 & yes & 0 \\
1.50 & $0$ & 1.94 & 0.9998 & yes & 0 \\
1.75 & $-2$ & 1.95 & 0.9998 & yes & 0 \\
2.50 & $0$ & 1.89 & 0.9998 & yes & 0 \\
\bottomrule
\end{tabular}
\end{table}

Using the matter-era criterion $|\omega_{eff}|<0.05$ and $\Omega_m>0.8$, the longest matter-era e-folds of every radiation-initialized orbit in Table~\ref{tab:radiation-runs} is zero, confirming that no identifiable matter-dominated segment forms.

To place a quantitative upper bound on how long matter domination can be maintained when an orbit is placed by hand near the matter point, 54 orbits were integrated starting from the $n=1$ saddle family with perturbations $\epsilon\in\{10^{-4},\allowbreak 10^{-3},\allowbreak 10^{-2}\}$ applied to the state components. Table~\ref{tab:neighbourhood} lists the six longest matter eras, the longest being 1.56 e-folds ($\Omega_m^{(0)}=0.90$, $\epsilon=10^{-4}$, perturbation of the $y$ component). Increasing $\epsilon$ or taking $\Omega_m^{(0)}\to1$ shortens the matter era, the latter in agreement with the divergence of $\lambda_+$ as $(1-\Omega_m)^{-1}$. Compared with the roughly 8.5 e-folds required observationally, 1.56 e-folds---achieved only with delicately hand-tuned initial data---falls far short of supporting structure formation \citep{delaurentis2015cosmological}.

\begin{table}[htbp]
\caption{The longest matter eras among the 54 saddle-neighbourhood orbits of Model~1.}
\label{tab:neighbourhood}
\small
\begin{tabular}{@{}cllll@{}}
\toprule
$\Omega_m^{(0)}$ & $\epsilon$ & Perturbed component & $N_{\rm end}$ & Matter-era e-folds \\
\midrule
0.900 & $10^{-4}$ & $y$ & 20.00 & 1.56 \\
0.900 & $10^{-4}$ & $x_3$ & 20.00 & 1.18 \\
0.900 & $10^{-3}$ & $y$ & 20.00 & 1.18 \\
0.950 & $10^{-4}$ & $y$ & 20.00 & 1.08 \\
0.950 & $10^{-3}$ & $y$ & 20.00 & 0.82 \\
0.900 & $10^{-2}$ & $y$ & 20.00 & 0.78 \\
\bottomrule
\end{tabular}
\end{table}

Taken together, Tables~\ref{tab:fp}--\ref{tab:neighbourhood} and Figs.~\ref{fig:model1-evolution} and \ref{fig:model1-phase} leave the verdict on Model~1 unambiguous within the closed system: the two end fixed points exist (the radiation point can act as an early transient and the de~Sitter point is a transversal attractor), but the intermediate continuous saddle family is doubly constrained by the closure singularity and the diverging positive eigenvalue, while at the trajectory level the orbits are cut off by the intrinsic $\mathcal{G}=0$ singularity; no complete heteroclinic evolution chain exists.

\subsection{Parameter regions and fixed-point spectra of Model 2}
\label{sec:results-model2}

Table~\ref{tab:lambda} gives the fourth eigenvalue $\lambda_{x_5}$ of each fixed point of Model~2 on the $x_5=0$ submanifold, with the corresponding results of the original model of Santos~da~Costa et al.\ ($m=1-n$) listed for comparison. Replacing the exponent constraint $1-n$ by $1-n/2$ systematically displaces the critical values of $n$: the zeros of $\lambda_{x_5}$ at $P_1$ and $P_5$ move from $n=1$ to $n=2$, at $P_8$ from $n=(2\pm\sqrt{6})/4$ to $n=1/2,3/2$, and at $P_2$ from $n=-2,1$ to $n=-2,2$.

\begin{table}[htbp]
\caption{The eigenvalue $\lambda_{x_5}$ of Model~2 and the new critical values of $n$.}
\label{tab:lambda}
\small
\begin{tabular}{@{}llll@{}}
\toprule
Point & $m=1-n$ \citep{dacosta2018dynamical} & $m=1-n/2$ (Model~2) & New critical $n$ \\
\midrule
$P_1$ & $8(n-1)/n$ & $4(n-2)/n$ & $2$ \\
$P_2$ & $-\dfrac{4(n-1)(n+2)}{2n^2-3n-1}$ & $-\dfrac{2(n-2)(n+2)}{2n^2-3n-1}$ & $-2,\ 2$ \\
$P_3$ & $4(2n-1)$ & $4(n-1)$ & $1$ \\
$P_4$ & $\dfrac{8n^2-12n+5}{n-1}$ & $\dfrac{4n^2-8n+5}{n-1}$ & no real root \\
$P_5$ & $6(n-1)/n$ & $3(n-2)/n$ & $2$ \\
$P_8$ & $\dfrac{8n^2-12n+3}{n-1}$ & $\dfrac{(2n-3)(2n-1)}{n-1}$ & $1/2,\ 3/2$ \\
\bottomrule
\end{tabular}
\end{table}

The finite scan can be replaced by an analytic classification on the full real line. Linearizing Eqs.~(38) and (39) of Ref.~\citep{dacosta2018dynamical} at $P_2$ gives the spectrum
\begin{equation}
\eqbox{\begin{aligned}
\lambda\in\Bigg\{&-\frac{2(n-2)(n+2)}{d},\ -\frac{2(5n^2-4n-2)}{d},\\
&-\frac{(n-1)(8n+3)}{d},\ -\frac{4(n+2)(n-1)}{d}\Bigg\},\quad d=2n^2-3n-1,
\end{aligned}}
\label{eq:p2-spectrum}
\end{equation}
which agrees with direct numerical Jacobians at representative points. Since $\omega_{eff}+1/3=-2(2n^2-2n+1)/(3d)$ and $2n^2-2n+1>0$, acceleration holds exactly for $d>0$, namely $n<n_-=(3-\sqrt{17})/4$ or $n>n_+=(3+\sqrt{17})/4$. Requiring all four eigenvalues to be negative then leaves only
\begin{equation}
\eqbox{n<-2\quad\text{or}\quad n>2,}
\label{eq:model2-range}
\end{equation}
with $n=\pm2$ excluded because $\lambda_{x_5}=0$ there (and the last eigenvalue also vanishes at $n=-2$). The finite scan therefore sampled two unbounded branches rather than fixing their boundaries.

For completeness, the full classification is as follows. For $-2<n<n_-$, $P_2$ is accelerating but unstable: a saddle on $-2<n<-3/8$ and $-3/8<n<(2-\sqrt{14})/5$, non-hyperbolic at $n=-3/8$ and $n=(2-\sqrt{14})/5$, and a repeller on $(2-\sqrt{14})/5<n<n_-$. For $n_-<n<n_+$, $d<0$ and the point is not accelerating; it is attracting on $n_-<n<1$ apart from the singular values $n=0$ and $(3-\sqrt{11})/4$, and a saddle on $1<n<n_+$, with $n=(2+\sqrt{14})/5$ non-hyperbolic and $(3+\sqrt{11})/4$ excluded by the source system. For $n_+<n<2$, the point accelerates but remains a saddle. At $n=n_\pm$ the denominator $d$ vanishes and $\omega_{eff}$ diverges. The excluded singular values of the underlying reduced system are $n=0,1,n_\pm,(3\pm\sqrt{11})/4$.

On the negative stable branch, $\omega_{eff}$ tends to $-1$ from above as $n\to-\infty$, rises to its least negative value $-0.96855$ at $n=-2-\sqrt{26}/2\simeq-4.5495$, and returns to $-1$ at the non-hyperbolic endpoint $n=-2$. On the positive stable branch it rises monotonically from $-11/3$ at $n\to2^+$ to $-1$ from below as $n\to+\infty$. For $n_+<n<2$, $\omega_{eff}$ rises from $-\infty$ toward $-11/3$ as $n\to2$; the interior extremum at $n=-2+\sqrt{26}/2\simeq0.5495$ lies in the non-accelerating central region. Compared with the original range $n<-1.59$ or $n>n_+$ \citep{dacosta2018dynamical}, the lower boundary moves to $n=-2$ because the zero of $\lambda_{x_5}(P_2)$ moves from $n=1$ to $n=2$; $n_+<n<2$ accelerates but is a saddle. The stable-accelerating branches therefore exist in the full domain, but neither supplies a standard matter-dominated stage.

\begin{figure}[htbp]
\centering
\includegraphics[width=\columnwidth]{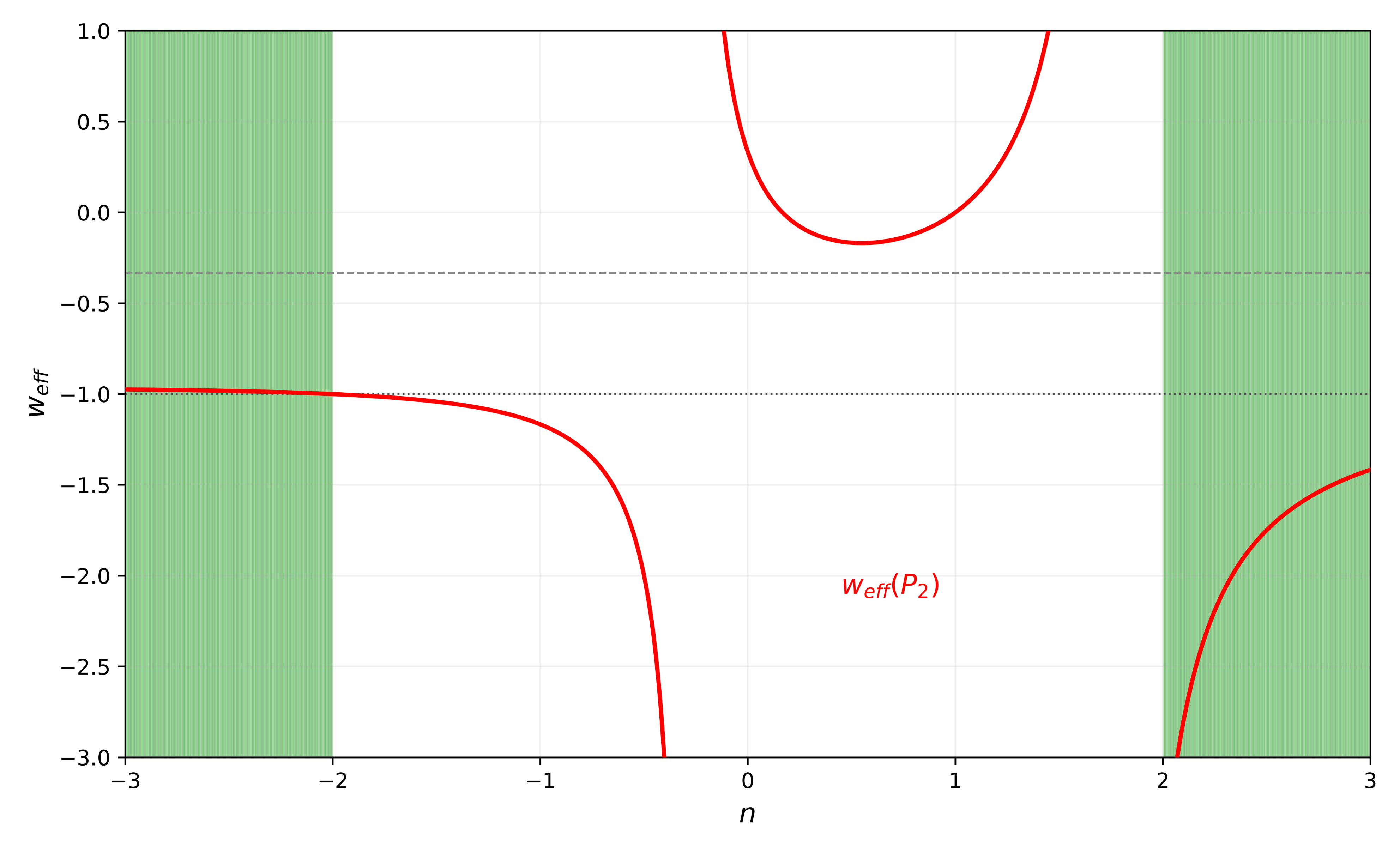}
\caption{Stability criteria of Model~2. Horizontal axis $n$, vertical axis $\omega_{eff}(P_2)$; the green shaded band shows the stable-accelerating intervals $n<-2$ and $n>2$ (the plotted portion lies in $-3<n<-2$ and $2<n<3$). The curve is given by $\omega_{eff}=-(6n^2-7n+1)/(6n^2-9n-3)$, and the stability boundaries follow from the analytic spectrum of Eqs.~(38) and (39) of Ref.~\citep{dacosta2018dynamical}.}
\label{fig:model2-stability}
\end{figure}

Figure~\ref{fig:model2-stability} plots the resulting $\omega_{eff}(P_2)$ and marks the stable-accelerating bands. Relative to Fig.~1 of Santos~da~Costa et al.\ \citep{dacosta2018dynamical}, the right-hand band starts at $n\simeq2$ rather than $1.78$.

Table~\ref{tab:stability} summarizes the stability classification and $\omega_{eff}(P_2)$ for representative values of $n$. The standard radiation point $P_3$ is a saddle for all listed values and can act as an early transient; $P_4$ is a ``wrong matter era'' with $\Omega_m=2$ and $\omega_{eff}=1/3$ throughout; $P_5$ becomes an attractor only in the vicinity of $0.6<n<0.75$, but there $\omega_{eff}>0$ and $\Omega_m$ is far greater than 1, so it does not constitute standard matter domination.

\begin{table}[htbp]
\caption{Stability classification and effective equation of state of Model~2 (direct numerical Jacobians).}
\label{tab:stability}
\small
\begin{tabular}{@{}lllllllc@{}}
\toprule
$n$ & $P_2$ & $P_3$ & $P_4$ & $P_5$ & $P_8$ & $\omega_{eff}(P_2)$ \\
\midrule
$-3$ & attractor & saddle & saddle & saddle & saddle & $-0.974$ \\
$-2$ & attractor & saddle & saddle & saddle & saddle & $-1.000$ \\
$-1$ & saddle & saddle & saddle & saddle & saddle & $-1.167$ \\
0.6 & attractor & saddle & saddle & attractor & repeller & $-0.167$ (not accelerating) \\
0.9 & attractor & saddle & attractor & saddle & repeller & $-0.071$ (not accelerating) \\
0.98 & attractor & saddle & attractor & saddle & repeller & $-0.016$ (not accelerating) \\
1.05 & saddle & saddle & saddle & saddle & saddle & $+0.045$ \\
1.50 & saddle & saddle & saddle & saddle & repeller & $+1.333$ \\
2.50 & attractor & saddle & saddle & saddle & repeller & $-1.750$ (accelerating) \\
3.50 & attractor & saddle & saddle & saddle & repeller & $-1.282$ (accelerating) \\
\bottomrule
\end{tabular}
\end{table}

Multi-start numerical root finding on the branches with $x_5\neq0$ gives: for $n=-2$ ($m=2$) there are 5 fixed points and the unique attractor is the de~Sitter solution with $\omega_{eff}=-1$ and $\Omega_m=0$; for $n=0.9$ ($m=0.55$) there are 8 fixed points, the two attractors being $\omega_{eff}=-0.0705$ and the ``wrong matter era'' with $\Omega_m=2,\omega_{eff}=1/3$, while the de~Sitter point ($x_3=2,x_5=-3.222$) is a saddle; for $n=3.5$ ($m=-0.75$) there are 8 fixed points and the unique attractor is the phantom-accelerating solution with $\omega_{eff}=-1.2821$ and $\Omega_m=0$. Under none of the three values of $n$ is a standard matter-dominated point with $\Omega_m\simeq1$ and $\omega_{eff}\simeq0$ found.

\begin{figure}[htbp]
\centering
\includegraphics[width=\columnwidth]{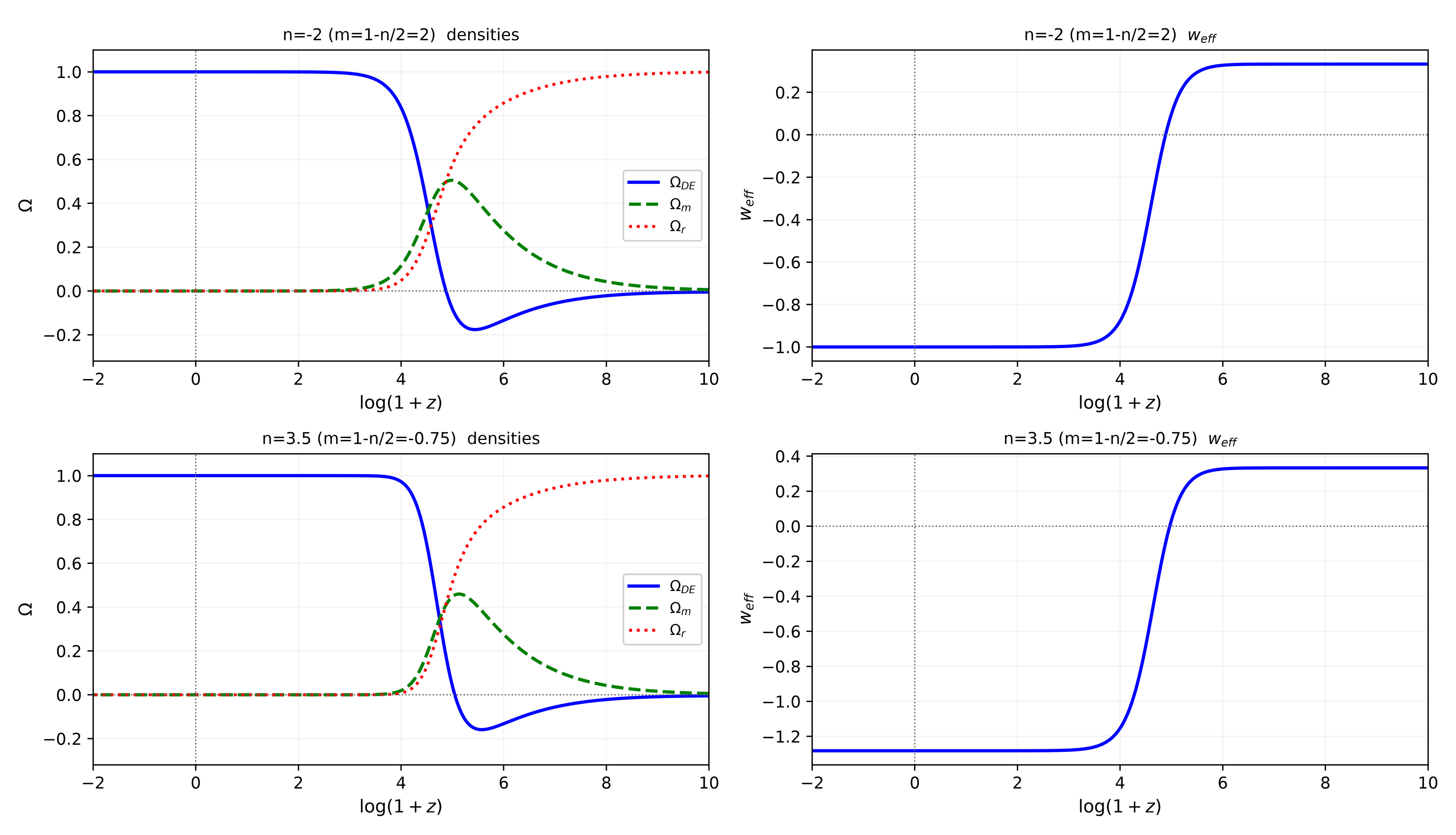}
\caption{Evolution trajectories of Model~2 within the reduced system of Eqs.~(38) and (39) of Ref.~\citep{dacosta2018dynamical}. Upper row $n=-2$, lower row $n=3.5$; left column $\Omega_{DE},\Omega_m,\Omega_r$, right column $\omega_{eff}$; horizontal axis $\log(1+z)$. Initial data $x_3=10^{-9}$, $x_4=0.999$, $x_5=0$, $x_7=0.0059$, $z=2.2\times10^{4}$. The matter-dominated segment is missing: radiation passes directly into acceleration.}
\label{fig:model2-evolution}
\end{figure}

The density and $\omega_{eff}$ evolution in Fig.~\ref{fig:model2-evolution} confirms the trajectory-level result: for both $n=-2$ and $n=3.5$, radiation passes directly to $\Omega_{DE}\to1$, with terminal $\omega_{eff}=-1$ and $-1.282$, respectively. The matter density shows only a brief bump, never a standard matter-dominated segment.

\section{Discussion}
\label{sec:discussion}

\subsection{Stage-by-stage verdict on the three-stage evolution chain}
\label{sec:discussion-verdict}

A complete chain requires a radiation fixed point with $\omega_{eff}=1/3$ and $\Omega_r\simeq1$, a saddle matter point with $\omega_{eff}=0$ and $\Omega_m\simeq1$, and a stable accelerating attractor with $\omega_{eff}<-1/3$ and $\Omega_{GDE}\simeq1$; a missing stage, misplaced stability, or inaccessible orbit invalidates the chain \citep{dong2019noether,dacosta2018dynamical}. Observational anchors are BBN and CMB measurements for the radiation era \citep{amendola2007conditions,hinshaw2013nine,ade2016planck,aghanim2020planck}, CMB acoustic peaks and $\Delta N\approx8.5$ e-folds of structure formation for matter domination \citep{amendola2007accelerating,delaurentis2015cosmological}, and supernovae, BAO, and the statefinder calibration $\Lambda$CDM: $r=1,s=0$ for acceleration \citep{alam2003exploring,riess1998observational,perlmutter1999measurements,cole20052dfgrs,eisenstein2005detection,aubourg2015cosmological,carvalho2016constraints}.

Against these criteria, the radiation end is available in both models: for Model~1 the $p=0$ radiation point acts as a repelling end when $n<1$, and all 18 radiation-initialized orbits start from $\Omega_r\simeq1$; for Model~2 $P_3$ is a standard radiation saddle with $\Omega_r=1$ \citep{dacosta2018dynamical}. The standard matter-dominated stage fails in both. In Model~1 the $n=1$ continuous saddle family gives $\omega_{eff}=0$, but $\Omega_m=1$ is a zero-over-zero limiting point of the closure, where $\lambda_+$ diverges only as $\Omega_m\to1^-$ from the physical side; the matter-era e-folds vanish for all 18 radiation-initialized orbits, and the longest of the 54 saddle-neighbourhood orbits is 1.56 e-folds, far below the $\simeq8.5$ required by structure formation \citep{delaurentis2015cosmological}. In Model~2, $\omega_{eff}\approx0$ occurs only at the excluded singular value $n=1$ for $P_2$ or in the $\Omega_m\to0$ limit of $P_5$ as $n\to1$; $P_4$ is the ``wrong matter era'' with $\Omega_m=2$ and $\omega_{eff}=1/3$ throughout, and the numerical root search with $x_5\neq0$ finds no standard matter point \citep{dacosta2018dynamical}.

The accelerating end is formally present but orbitally inaccessible in Model~1: the de~Sitter spectrum $\{-4,-3,-3,0\}$ is transversally attracting, yet none of the 18 radiation-initialized orbits reaches it and 12 reach $\mathcal{G}=0$. In Model~2, $P_2$ is stable and accelerating for $n<-2$ or $n>2$, with an end state $\Omega_{DE}\to1$ and $\omega_{eff}=-1$ for $n=-2$ or $-1.282$ for $n=3.5$. A complete heteroclinic chain would require the three stages to connect sequentially for the same parameters with accessible orbits \citep{dacosta2018dynamical,delaurentis2015cosmological}. Neither model meets this requirement, and none of the three source works performs an observational data fit \citep{dong2019noether,dacosta2018dynamical,singh2025dynamical}. Both are therefore judged ``no'': Model~1 within the closed system and Model~2 as a formal extrapolation within the reduced system \citep{dong2019noether}.

\subsection{Comparison with existing dynamical-systems results}
\label{sec:discussion-comparison}

For the $m=1-n$ branch, Santos~da~Costa et al.\ found eight fixed points and no complete radiation-to-acceleration trajectory; their stable-accelerating interval for $P_2$ was $n<-1.59$ or $n>(3+\sqrt{17})/4\approx1.7808$ \citep{dacosta2018dynamical}. Substituting $m=1-n/2$ shifts this interval to $n<-2$ or $n>2$ because the zero of $\lambda_{x_5}(P_2)$ moves from $n=1$ to $n=2$ (Table~\ref{tab:lambda}).

Singh \citep{singh2025dynamical} studied the $m=1-\alpha$ and $m=1$ cases, each with nine fixed points, and listed observational tests as future work without using data sets. Model~1 meets the $m=1$ line only at $n=2$, an open boundary of the $Q_1$ and $Q_9$ stability ranges where linearization degenerates; in the $(n,m)$ plane its intersections with $m=1-n$ at $(2/3,1/3)$ and with $m=1$ at $(2,1)$ are likewise degenerate or critical. Model~2 meets $m=1-n$ only at $(0,1)$, where $f=k\mathcal{G}$ is topological and $\alpha=0$ is excluded \citep{dong2019noether,dacosta2018dynamical,singh2025dynamical}. All three analyses use Jacobian eigenvalue signs and find radiation saddles and de~Sitter attractors but no standard matter era; they differ in exponent relations, critical $n$, and fixed-point counting, none matching the continuous family $\mathcal{M}_{\rm vac}$ of closed Model~1 \citep{dacosta2018dynamical,singh2025dynamical}.

\subsection{Methodological boundaries and untested items}
\label{sec:discussion-boundaries}

The Model~2 conclusion is confined to the reduced system of Ref.~\citep{dacosta2018dynamical}: $m$ was kept general and set to $1-n$ only when solving for the fixed points, so inserting $1-n/2$ is legitimate at that level \citep{dong2019noether,dacosta2018dynamical}. The extrapolation inherits three structural premises---the dropped $f_{R\mathcal{G}}\dot{\mathcal{G}}$ term, the unimposed identity $x_5=mx_2=(m/n)x_3$, and the sign convention of Eq.~(36) of Ref.~\citep{dacosta2018dynamical}---and is not a strict verdict on Eq.~(67) of Ref.~\citep{dong2019noether}.

For Model~1, promoting $\dot R$ gives an explicit $\Gamma(x_i)$; the restriction of the choice of state variable is retained: a non-power-law $f(\mathcal{G})$ or a regularizing term as $\mathcal{G}\to0$ would alter trajectory accessibility \citep{dong2019noether,dacosta2018dynamical}. None of the three source works fits observational data, $n$ is not observationally fixed, and ghost-mode and scalar-perturbation stability remain untested \citep{dong2019noether,dacosta2018dynamical,singh2025dynamical,defelice2010matter,defelice2010final,calcagni2006ghost}.

\subsection{Feasible directions of improvement}
\label{sec:discussion-improvements}

The improvement paths reduce to three directions. First, the $\mathcal{G}$ dependence should be changed to remove the $\mathcal{G}=0$ singularity, e.g.\ by adopting a non-power-law $f(\mathcal{G})$ or adding a regularizing term as $\mathcal{G}\to0$. Second, the matter sector should be decoupled from the geometric dark energy so that the standard matter-dominated era is no longer tied to the singularity of the geometric term. Third, $n$ should be constrained with observational data from type~Ia supernovae, baryon acoustic oscillations, and the CMB \citep{dong2019noether,riess1998observational,perlmutter1999measurements,hinshaw2013nine,cole20052dfgrs,eisenstein2005detection,delaurentis2015cosmological}. The dimension-raising closure of Model~1 is not restricted to the power-law form with $m=n/2$ and can be generalized to other $f(R,\mathcal{G})$ models to reconstruct explicitly $n$-dependent Jacobian spectra \citep{dong2019noether,dacosta2018dynamical}. For Model~2, the reduced system must be rebuilt after correcting the $f_{R\mathcal{G}}\dot{\mathcal{G}}$ term and imposing the identity $x_5=mx_2=(m/n)x_3$, and only then can a dynamical verdict belonging to Eq.~(67) of Ref.~\citep{dong2019noether} itself be given for the fixed-point spectrum and the stability intervals \citep{dong2019noether,dacosta2018dynamical}.

\section{Conclusions}
\label{sec:conclusion}

We have tested separately whether the two generalized-Noether-selected $f(R,\mathcal{G})$ models, with $m=n/2$ and $m=1-n/2$, can realize radiation domination, standard matter domination, and late-time acceleration in sequence. Model~1 was analysed in a closed four-dimensional autonomous system. Its vacuum family has the explicit spectrum $\{0,\ 2n(2-p)-4,\ 2n(2-p)-3,\ (4n-5)+(1-2n)p\}$, verified to $6.98\times10^{-5}$ over 30 parameter sets, and the de~Sitter point has the $n$-independent spectrum $\{-4,-3,-3,0\}$. Although an $\omega_{eff}=0$ saddle family exists at $n=1$, $\Omega_m=1$ is a zero-over-zero limiting point of the closure: $D=0$ and $A_0-2\delta=0$, so the present formulation does not define the pure-matter end point; the positive eigenvalue diverges only as $\Omega_m\to1^-$ from the physical side. None of 18 radiation-initialized orbits reaches de~Sitter, 12 reach $\mathcal{G}=0$, and all matter-era e-folds vanish; the longest matter era among 54 saddle-neighbourhood orbits is only 1.56 e-folds. Within the chosen closure, the $\mathcal{G}=0$ barrier separates the radiation and de~Sitter regions.

For Model~2, the formal reduced-system extrapolation shifts the stable-accelerating interval of $P_2$ to $n<-2$ or $n>2$, but no standard matter-dominated point is found and the evolution passes directly from radiation to acceleration. This verdict inherits the premises of the reduced system and is not strict for Eq.~(67) of Ref.~\citep{dong2019noether}.

Both models are therefore judged ``no'' within their respective verifiable scopes. No observational fit is claimed, and the allowed range of $n$, ghost modes, and scalar perturbations remain open. A viable history would require a regularized $\mathcal{G}$ dependence, separation of the matter sector from geometric dark energy, and observational constraints on $n$; Model~2 additionally requires rebuilding the reduced system with the cross-derivative term and the phase-space identity restored.

\bibliography{arxiv_paper}

@article{dong2019noether,
  author  = {Dong, Han},
  title   = {Generalized Noether theorem for Gauss--Bonnet cosmology},
  journal = {Chin. J. Phys.},
  volume  = {60},
  pages   = {573},
  year    = {2019},
  doi     = {10.1016/j.cjph.2019.06.001}
}

@article{dacosta2018dynamical,
  author  = {Santos da Costa, S. and Roig, F. V. and Alcaniz, J. S. and Capozziello, S. and De Laurentis, M. and Benetti, M.},
  title   = {Dynamical analysis on $f(R,\mathcal{G})$ cosmology},
  journal = {Class. Quantum Grav.},
  volume  = {35},
  pages   = {075013},
  year    = {2018},
  doi     = {10.1088/1361-6382/aaad80}
}

@article{singh2025dynamical,
  author  = {Singh, A.},
  title   = {Dynamical systems of modified Gauss--Bonnet gravity: cosmological implications},
  journal = {Eur. Phys. J. C},
  volume  = {85},
  pages   = {24},
  year    = {2025},
  doi     = {10.1140/epjc/s10052-024-13732-3}
}

@article{capozziello2014noether,
  author  = {Capozziello, S. and De Laurentis, M. and Odintsov, S. D.},
  title   = {Noether symmetry approach in Gauss--Bonnet cosmology},
  journal = {Mod. Phys. Lett. A},
  volume  = {29},
  pages   = {1450164},
  year    = {2014},
  doi     = {10.1142/S0217732314501648}
}

@article{amendola2007conditions,
  author  = {Amendola, L. and Gannouji, R. and Polarski, D. and Tsujikawa, S.},
  title   = {Conditions for the cosmological viability of $f(R)$ dark energy models},
  journal = {Phys. Rev. D},
  volume  = {75},
  pages   = {083504},
  year    = {2007},
  doi     = {10.1103/PhysRevD.75.083504}
}

@article{riess1998observational,
  author  = {Riess, A. G. and others},
  title   = {Observational evidence from supernovae for an accelerating universe and a cosmological constant},
  journal = {Astron. J.},
  volume  = {116},
  pages   = {1009},
  year    = {1998},
  doi     = {10.1086/300499}
}

@article{perlmutter1999measurements,
  author  = {Perlmutter, S. and others},
  title   = {Measurements of $\Omega$ and $\Lambda$ from 42 high-redshift supernovae},
  journal = {Astrophys. J.},
  volume  = {517},
  pages   = {565},
  year    = {1999},
  doi     = {10.1086/307221}
}

@article{hinshaw2013nine,
  author  = {Hinshaw, G. and others},
  title   = {Nine-year Wilkinson Microwave Anisotropy Probe ({WMAP}) observations: cosmological parameter results},
  journal = {Astrophys. J. Suppl.},
  volume  = {208},
  pages   = {19},
  year    = {2013},
  doi     = {10.1088/0067-0049/208/2/19}
}

@article{ade2016planck,
  author  = {Ade, P. A. R. and others},
  title   = {Planck 2015 results. {XIII}. Cosmological parameters},
  journal = {Astron. Astrophys.},
  volume  = {594},
  pages   = {A13},
  year    = {2016},
  doi     = {10.1051/0004-6361/201525830}
}

@article{aghanim2020planck,
  author  = {Aghanim, N. and others},
  title   = {Planck 2018 results. {VI}. Cosmological parameters},
  journal = {Astron. Astrophys.},
  volume  = {641},
  pages   = {A6},
  year    = {2020},
  doi     = {10.1051/0004-6361/201833910}
}

@article{cole20052dfgrs,
  author  = {Cole, S. and others},
  title   = {The 2dF Galaxy Redshift Survey: power-spectrum analysis of the final data set and cosmological implications},
  journal = {Mon. Not. R. Astron. Soc.},
  volume  = {362},
  pages   = {505},
  year    = {2005},
  doi     = {10.1111/j.1365-2966.2005.09318.x}
}

@article{eisenstein2005detection,
  author  = {Eisenstein, D. J. and others},
  title   = {Detection of the baryon acoustic peak in the large-scale correlation function of {SDSS} luminous red galaxies},
  journal = {Astrophys. J.},
  volume  = {633},
  pages   = {560},
  year    = {2005},
  doi     = {10.1086/466512}
}

@article{weinberg1989cosmological,
  author  = {Weinberg, S.},
  title   = {The cosmological constant problem},
  journal = {Rev. Mod. Phys.},
  volume  = {61},
  pages   = {1},
  year    = {1989},
  doi     = {10.1103/RevModPhys.61.1}
}

@article{defelice2010f,
  author  = {De Felice, A. and Tsujikawa, S.},
  title   = {$f(R)$ theories},
  journal = {Living Rev. Rel.},
  volume  = {13},
  pages   = {3},
  year    = {2010},
  doi     = {10.12942/lrr-2010-3}
}

@article{nojiri2005gauss,
  author  = {Nojiri, S. and Odintsov, S. D.},
  title   = {Modified Gauss--Bonnet theory as gravitational alternative for dark energy},
  journal = {Phys. Lett. B},
  volume  = {631},
  pages   = {1},
  year    = {2005},
  doi     = {10.1016/j.physletb.2005.10.010}
}

@article{li2007cosmology,
  author  = {Li, B. and Barrow, J. D. and Mota, D. F.},
  title   = {Cosmology of modified Gauss--Bonnet gravity},
  journal = {Phys. Rev. D},
  volume  = {76},
  pages   = {044027},
  year    = {2007},
  doi     = {10.1103/PhysRevD.76.044027}
}

@article{defelice2009solar,
  author  = {De Felice, A. and Tsujikawa, S.},
  title   = {Construction of cosmologically viable $f(\mathcal{G})$ dark energy models},
  journal = {Phys. Lett. B},
  volume  = {675},
  pages   = {1},
  year    = {2009},
  doi     = {10.1016/j.physletb.2009.03.060}
}

@article{ivanov2012dynamics,
  author  = {Ivanov, M. M. and Toporensky, A. V.},
  title   = {Cosmological dynamics of fourth-order gravity with a Gauss--Bonnet term},
  journal = {Gravitation and Cosmology},
  volume  = {18},
  pages   = {43},
  year    = {2012},
  doi     = {10.1134/S0202289312010100}
}

@article{alimohammadi2009phase,
  author  = {Alimohammadi, M. and Ghalee, A.},
  title   = {Phase space of generalized Gauss--Bonnet dark energy},
  journal = {Phys. Rev. D},
  volume  = {80},
  pages   = {043006},
  year    = {2009},
  doi     = {10.1103/PhysRevD.80.043006}
}

@article{nojiri2005dark,
  author  = {Nojiri, S. and Odintsov, S. D. and Sasaki, M.},
  title   = {Gauss--Bonnet dark energy},
  journal = {Phys. Rev. D},
  volume  = {71},
  pages   = {123509},
  year    = {2005},
  doi     = {10.1103/PhysRevD.71.123509}
}

@article{delaCruzDombriz2012cosmological,
  author  = {de la Cruz-Dombriz, A. and S{\'a}ez-G{\'o}mez, D.},
  title   = {On the stability of the cosmological solutions in $f(R,\mathcal{G})$ gravity},
  journal = {Class. Quantum Grav.},
  volume  = {29},
  pages   = {245014},
  year    = {2012},
  doi     = {10.1088/0264-9381/29/24/245014}
}

@article{amendola2007accelerating,
  author  = {Amendola, L. and Polarski, D. and Tsujikawa, S.},
  title   = {Are $f(R)$ dark energy models cosmologically viable?},
  journal = {Phys. Rev. Lett.},
  volume  = {98},
  pages   = {131302},
  year    = {2007},
  doi     = {10.1103/PhysRevLett.98.131302}
}

@article{delaurentis2015cosmological,
  author  = {De Laurentis, M. and Paolella, M. and Capozziello, S.},
  title   = {Cosmological inflation in $f(R,\mathcal{G})$ gravity},
  journal = {Phys. Rev. D},
  volume  = {91},
  pages   = {083531},
  year    = {2015},
  doi     = {10.1103/PhysRevD.91.083531}
}

@article{carloni2005dynamical,
  author  = {Carloni, S. and Dunsby, P. K. S. and Capozziello, S. and Troisi, A.},
  title   = {Cosmological dynamics of $R^n$ gravity},
  journal = {Class. Quantum Grav.},
  volume  = {22},
  pages   = {4839},
  year    = {2005},
  doi     = {10.1088/0264-9381/22/22/011}
}

@article{carloni2015dynamical,
  author  = {Carloni, S. and Koivisto, T. and Lobo, F. S. N.},
  title   = {Dynamical system analysis of hybrid metric-{Palatini} cosmologies},
  journal = {Phys. Rev. D},
  volume  = {92},
  pages   = {064035},
  year    = {2015},
  doi     = {10.1103/PhysRevD.92.064035}
}

@book{birrell1982quantum,
  author    = {Birrell, N. D. and Davies, P. C. W.},
  title     = {Quantum Fields in Curved Space},
  publisher = {Cambridge University Press},
  address   = {Cambridge},
  year      = {1982},
  doi       = {10.1017/CBO9780511622632}
}

@article{bajardi2023dynamical,
  author  = {Bajardi, F. and D'Agostino, R.},
  title   = {Late-time constraints on modified Gauss--Bonnet cosmology},
  journal = {Gen. Relativ. Gravit.},
  volume  = {55},
  pages   = {49},
  year    = {2023},
  doi     = {10.1007/s10714-023-03092-w}
}

@article{alam2003exploring,
  author  = {Alam, U. and Sahni, V. and Saini, T. D. and Starobinsky, A. A.},
  title   = {Exploring the expanding Universe and dark energy using the statefinder diagnostic},
  journal = {Mon. Not. R. Astron. Soc.},
  volume  = {344},
  pages   = {1057},
  year    = {2003},
  doi     = {10.1046/j.1365-8711.2003.06871.x}
}

@article{aubourg2015cosmological,
  author  = {Aubourg, {\'E}. and others},
  title   = {Cosmological implications of baryon acoustic oscillation measurements},
  journal = {Phys. Rev. D},
  volume  = {92},
  pages   = {123516},
  year    = {2015},
  doi     = {10.1103/PhysRevD.92.123516}
}

@article{defelice2010matter,
  author  = {De Felice, A. and G{\'e}rard, J.-M. and Suyama, T.},
  title   = {Cosmological perturbation in $f(R,\mathcal{G})$ theories with a perfect fluid},
  journal = {Phys. Rev. D},
  volume  = {82},
  pages   = {063526},
  year    = {2010},
  doi     = {10.1103/PhysRevD.82.063526}
}

@book{olver1993applications,
  author    = {Olver, P. J.},
  title     = {Applications of Lie Groups to Differential Equations},
  publisher = {Springer},
  address   = {New York},
  year      = {1993},
  doi       = {10.1007/978-1-4612-4350-2}
}

@article{dong2013new,
  author  = {Dong, Han and Wang, Jiaxin and Meng, Xinhe},
  title   = {The distinctions between $\Lambda$CDM and $f(T)$ gravity according to Noether symmetry},
  journal = {Eur. Phys. J. C},
  volume  = {73},
  pages   = {2543},
  year    = {2013},
  doi     = {10.1140/epjc/s10052-013-2543-2}
}

@article{capozziello2016noether,
  author  = {Capozziello, S. and De Laurentis, M. and Dialektopoulos, K. F.},
  title   = {Noether symmetries in Gauss--Bonnet-teleparallel cosmology},
  journal = {Eur. Phys. J. C},
  volume  = {76},
  pages   = {629},
  year    = {2016},
  doi     = {10.1140/epjc/s10052-016-4491-0}
}

@article{cognola2006dark,
  author  = {Cognola, G. and Elizalde, E. and Nojiri, S. and Odintsov, S. D. and Zerbini, S.},
  title   = {Dark energy in modified Gauss--Bonnet gravity: late-time acceleration and the hierarchy problem},
  journal = {Phys. Rev. D},
  volume  = {73},
  pages   = {084007},
  year    = {2006},
  doi     = {10.1103/PhysRevD.73.084007}
}

@article{defelice2010final,
  author  = {De Felice, A. and Tanaka, T.},
  title   = {Inevitable ghost and the degrees of freedom in $f(R,\mathcal{G})$ gravity},
  journal = {Prog. Theor. Phys.},
  volume  = {124},
  pages   = {503},
  year    = {2010},
  doi     = {10.1143/PTP.124.503}
}

@article{calcagni2006ghost,
  author  = {Calcagni, G. and de Carlos, B. and De Felice, A.},
  title   = {Ghost conditions for Gauss--Bonnet cosmologies},
  journal = {Nucl. Phys. B},
  volume  = {752},
  pages   = {404},
  year    = {2006},
  doi     = {10.1016/j.nuclphysb.2006.06.020}
}

@article{krauss1995age,
  author  = {Krauss, L. M. and Turner, M. S.},
  title   = {The cosmological constant is back},
  journal = {Gen. Rel. Grav.},
  volume  = {27},
  pages   = {1137},
  year    = {1995},
  doi     = {10.1007/BF02108229}
}

@article{alcaniz1999cosmological,
  author  = {Alcaniz, J. S. and Lima, J. A. S.},
  title   = {New limits on $\Omega_\Lambda$ and $\Omega_M$ from old galaxies at high redshift},
  journal = {Astrophys. J.},
  volume  = {521},
  pages   = {87},
  year    = {1999},
  doi     = {10.1086/312191}
}

@article{alcaniz2001constraining,
  author  = {Alcaniz, J. S. and Lima, J. A. S.},
  title   = {Dark energy and the epoch of galaxy formation},
  journal = {Astrophys. J.},
  volume  = {550},
  pages   = {133},
  year    = {2001},
  doi     = {10.1086/319642}
}

@article{carvalho2016constraints,
  author  = {Carvalho, G. C. and others},
  title   = {Baryon acoustic oscillations from the {SDSS DR10} galaxies angular correlation function},
  journal = {Phys. Rev. D},
  volume  = {93},
  pages   = {023530},
  year    = {2016},
  doi     = {10.1103/PhysRevD.93.023530}
}

\end{document}